# Molecular Dynamics Simulations of Heat Conduction in Nanostructures: Effect of Heat Bath

Jie Chen,[1] Gang Zhang,[2,*] and Baowen Li[1,3,&]

[1]Department of Physics and Centre for Computational Science and Engineering, National University of Singapore, Singapore 117546, Singapore

[2]Department of Electronics, Peking University, Beijing 100871, People's Republic of China

[3]NUS Graduate School for Integrative Sciences and Engineering, Singapore 117456, Singapore

## ABSTRACT

We investigate systematically the impacts of heat bath used in molecular dynamics simulations on heat conduction in nanostructures exemplified by Silicon Nanowires (SiNWs) and Silicon/Germanium nano junction. It is found that multiple layers of Nosé-Hoover heat bath are required to reduce the temperature jump at the boundary, while only a single layer of Langevin heat bath is sufficient to generate a linear temperature profile with small boundary temperature jump. Moreover, an intermediate value of heat bath parameter is recommended for both Nosé-Hoover and Langevin heat bath in order to achieve correct temperature profile and thermal conductivity in homogeneous materials. Furthermore, the thermal rectification ratio in Si/Ge thermal diode depends on the choice of Nosé-Hoover heat bath parameter remarkably, which may lead to non-physical results. In contrast, Langevin heat bath is recommended because it can produce consistent results with experiment in large heat bath parameter range.

[*] Email: zhanggang@pku.edu.cn
[&] Email: phylibw@nus.edu.sg

I. INTRODUCTION

Heat conduction in nanostructures is of great importance both from fundamental and application point of view. On the one hand, superior thermal conductivity has been observed in graphene [1, 2] and carbon nanotube [3] which has raised an exciting prospect of application for thermal devices. [4-8] On the other hand, in some nano materials such as silicon naowires (SiNWs), due to the strong surface inelastic scatterings, the thermal conductivity of SiNWs is about two orders of magnitude smaller than that of bulk crystals. [9] The low thermal conductivity of SiNWs is of particular interest for thermoelectric application. [10-12] Furthermore, the physical law works in bulk material such as Fourier law of heat conduction is not valid anymore in nanoscale [13-16]. Although there has been some progress in thermal conductivity research in recent years, it is still a great challenge to investigate the thermal property of nano materials experimentally due to the small size. Thus in order to understand the heat conduction in nanoscale, people depends heavily on computer simulation. Atomistic simulations, such as non-equilibrium molecular dynamics (NEMD) calculation plays an important role in thermal conductivity investigation since it can be used to study individual nano materials with realistic crystalline structures. Although some works about the heat transport in nanostructure have been carried out by using NEMD simulations, [14-18] there is still a lack of systematic understanding of the impacts of heat bath on the calculated thermal properties. In this paper, we provide a detailed study on the impacts of heat bath used in NEMD simulations on the predicted thermal conductivity, heat flux and thermal rectification ratio.

II. COMPUTATIONAL METHODS

In NEMD simulations, heat bath is used to set up temperature gradient in the system. There are two representative approaches used to control the temperature:

Nosé-Hoover (NH) heat bath [19, 20] which is an example of deterministic heat bath, and Langevin heat bath [21] which is an example of stochastic heat bath. The evolution of the particles in thermal contact with NH heat bath can be ruled by the equation as:

$$\frac{d\vec{q}_i}{dt} = \frac{\partial H}{\partial \vec{p}_i}, \qquad \frac{d\vec{p}_i}{dt} = -\frac{\partial H}{\partial \vec{q}_i} - \zeta \vec{p}_i, \qquad (1)$$

where $H$ is the Hamiltonian of the system, $\vec{p}_i$ and $\vec{q}_i$ are the momentum and coordinate of particle $i$, respectively, and $\zeta$ is an auxiliary variable modeling the microscopic action of the heat bath. The dynamics of $\zeta$ is governed by the following equation:

$$\frac{d\zeta}{dt} = \frac{1}{\tau^2}\left[\frac{\sum_{i\in S}\vec{p}_i \bullet \vec{p}_i}{3m_i k_B T N} - 1\right], \qquad (2)$$

where $k_B$ is the Boltzmann constant, $m_i$ is the mass of particle $i$, $T$ and $\tau$ are the aimed temperature and response time of heat bath, respectively, and $N$ is the total number of particles that is in contact with heat bath.

With Langevin heat bath, the equation of motion can be described as

$$\frac{d\vec{p}_i}{dt} = -\frac{\partial H}{\partial \vec{q}_i} + \xi - \lambda \vec{p}_i , \qquad (3)$$

where λ is the dissipation rate, and ξ is the random force that follows Wiener process with zero mean and variance $2m\lambda k_B T$ according to the fluctuation-dissipation theorem.

In both schemes of heat bathes, there is a free parameter, namely τ in NH heat bath and λ in Langevin heat bath, which controls the coupling between the system under study and heat bath. In this paper, we take SiNWs as an example to study the impact of heat bath on the calculated thermal properties of homogeneous materials.

Moreover, we extend our study to heterogeneous materials, such as Si/Ge NW junctions, in which a rectification of heat current in different directions is expected. To derive the force term, Stillinger-Weber (SW) potential [22, 23] is used. For the interaction between Si and Ge atoms, the net length and energy units in SW potential are taken to be the arithmetic average and geometric average of that of Si and Ge, respectively [18, 24]. The temperature of hot and cold heat baths are set as 310K and 290K, respectively. The simulations are performed long enough to allow the system to reach a non-equilibrium steady state where the heat current going through the system is time independent. All results given in this paper are obtained by averaging about $5\times10^7$ time steps, and each time step is set as 0.8 fs. Free boundary condition is used to atoms on the outer surface of the NWs. The thermal conductivity is calculated from the Fourier's law, $\kappa = -J_L/\nabla T$, where the heat current $J_L$ along the longitudinal direction is defined as the energy transported along the NW in unit time through the unit cross-section area, and $\nabla T$ is the temperature gradient.

III. RESULTS AND DISCUSSION

We first study the thermal conductivity of (100) SiNWs with a cross section of *3×3* unit cells (lattice constant is *0.543* nm, 8 atoms in each unit cell) and 10 unit cells in the longitudinal direction. The schematic diagram of SiNWs is shown in Fig. 1. Here we set longitudinal direction along *x* axis, and atoms in the same layers means they have the same *x* coordinate. At two ends of the NWs, fixed boundary condition is imposed on the boundary layers (pinpointed by arrows in Fig. 1). Next to the boundary layers, certain layers of SiNWs are put in contact with the heat bath (inside the rectangular box in Fig. 1). It has been reported that due to the lack of stability of the NW surfaces, it can lead to scattered computed results. [25] In our calculations, at room temperature and with the small temperature difference between the two ends, the surface structure is stable, and the different heat bath and the difference in temperature profile have no impact on the stability of NW surfaces. In Fig. 2 we show the effect of the number of heat bath layers on the thermal properties of SiNWs. Here $\tau = 0.1$ and $\lambda = 10$ are used in NH heat bath and Langevin heat bath, respectively. A linear temperature gradient is always observed in the interior, and the main difference in temperature profile between different heat bath conditions is the temperature jump between the heat bath layers and interior layers. With only one layer of NH heat bath, there exists a large temperature jump (TJ) between the heat bath layer and its neighbouring layer (shown in Fig. 2(a)), while the temperature jump is much smaller with one layer of Langevin heat bath (shown in Fig. 2(b)). This temperature jump can be explained by localized edge mode (LEM) of phonons. With fixed boundary condition, there exists edge mode localized at the neighbouring layer next to the fixed boundary [26]. This edge mode is actually a quite generic feature of materials in thermal transport and essentially originated from the specific geometrical configuration of the edge region [27], very similar to the electronic and/or spin edge states [28]. It should be emphasized that localization effect is a quite generic consequence of the broken of spatial periodicity due to the imposed boundary condition for finite system. Therefore, LEM also exists with other boundary conditions (e.g. free or periodic) [26]. Due to the localization, it does not contribute to heat transport. If heat bath is applied to this region, LEM will be excited and localized at this region, while other modes can propagate and contribute to the heat transport. We can see from Eq. (1) and (2) that in order to maintain a constant temperature, the mechanism of NH heat bath is to introduce a viscosity force which is proportional to the velocity, and the proportionality is determined from the velocities of all the particles. Due to this deterministic characteristic of NH heat bath, once LEM is excited, it will accumulate at the heat bath layer over time, because LEMs have a larger amplitude than other modes and account for a large percentage of the total modes [26]. As a result, although the heat bath can maintain a constant temperature, the major contribution comes from LEM which cannot be utilized in heat transport. Therefore, there exists a large TJ when there is only one layer of NH heat bath. With the number of heat bath layer increasing, the TJ will be reduced as shown in Fig. 2(a). This is because of the exponential decay feature of LEM over distance. With multiple layers of heat bath which are away from the localized region, other modes can be excited and dominate in heat transport. Therefore, the localized effect on the thermal transport is suppressed. The reduction in TJ leads to large increase in $J_L$ from NL=1 to NL=2 as shown in Fig. 2(d). Further increase of heat bath layers cannot eventually eliminate TJ. The small retained TJ is due to the thermal interface resistance between heat bath and rest parts.

With Langevin heat bath, the LEM can be get rid of automatically and result in a small TJ even with one layer of heat bath as shown in Fig. 2(b). In addition, when NL is increased, $J_L$ converges to a constant value much faster than the case with NH heat bath. The reason is that with Langevin heat bath, each mode can be excited randomly at every time step due to the stochastic characteristic of Langevin heat bath. Therefore, the accumulation effect of LEM over time will be suppressed and thus lead to a small TJ.

The large TJ induced by LEM can be further understood by looking at the autocorrelation function of velocity. Fig. 2(f) plots the normalized autocorrelation function of velocity of atoms in the middle of NWs with the same parameters used in Fig. 2(a) and (b). With one layer of NH heat bath, since it mainly utilizes LEM to maintain the constant temperature, there exists obviously nonvanishing correlation in the long-time region. This artificial correlation is the direct evidence of the existence and accumulation effect of LEM mentioned above, which leads to the large TJ. With multiple layers of NH, the artificial correlation vanishes because all the modes can contribute to the thermal transport, thus a much smaller TJ. Moreover, due to the stochastic characteristic of Langevin heat bath, the artificial correlation doesn't exist, regardless the number of heat bath layers applied.

It is worth mentioning that we have also checked the temperature jump in other popularly used deterministic heat bath, such as the Berendsen heat bath [29]. It is a velocity-scaling type heat bath, with the scaling factor [29]:

$$\gamma = \left[1 + \frac{\Delta t}{\tau_T}\left(\frac{T_0}{T} - 1\right)\right]^{1/2}, \tag{4}$$

where $\Delta t$ is the time step, $\tau_T$ is the relaxation time, $T_0$ and $T$ are the aimed temperature and instantaneous temperature, respectively. The relaxation time $\tau_T$ should be properly chosen to avoid unrealistically low temperature fluctuations with small $\tau_T$, and the inactive sampling with large $\tau_T$ (e.g. $\tau_T \to \infty$) [30]. Based on such considerations, we set $\tau_T = \Delta t * 10^4$ (8ps) in our study. As shown in Fig. 2(c), with Berendsen heat bath, the NW fails to reach the aimed temperature (310K and 290K) at two ends. More seriously, large temperature jump persists regardless the number of heat bath layers applied. This even poorer performance of Berendsen heat bath is an expected consequence of its inability of reproducing canonical ensemble, because of the artificial velocity-scaling scheme [30], which has been reported to cause artifacts in various studies [31]. Therefore, in the follows, we mainly concentrate on the parameter effect of both NH and Langevin heat bath.

From the calculated heat flux and temperature gradient, we can obtain the thermal conductivity. From Fig. 2(e), it is obvious that with Langevin heat bath, one can obtain a consistent result regardless of the number of heat bath layers applied. However, due to the existence of LEM and the deterministic nature of NH heat bath, multiple heat bath layers are required in order to get a consistent result. In both NH and Langevin heat bath, the thermal conductivity converges to a constant when $NL \geq 3$. However, the converged values of thermal conductivity are different in these two heat baths. Moreover, there is a free heat bath parameter ($\tau/\lambda$) which effectively controls the strength of noise (response time of NH heat bath/dissipate rate of Langevin heat bath). Since we fix the heat bath parameter in the first part of this study, whether the choice of free parameter in different heat bath is equivalent to each other and how it can influence the calculated thermal properties are yet not clear. In the following part, we fix the heat bath temperature at 310K and 290K, and tune the heat bath parameters to study their impacts on heat flux, temperature profile and thermal conductivity.

Next we set NL=3 for both NH and Langevin heat bath. In Fig. 3 we plot the impacts of heat bath parameter on the calculated thermal properties of SiNWs. For NH heat bath, τ cannot be too small (e.g. $\tau = 0.01$ in Fig. 3(a)) since it produces a wrong temperature profile in this case, although the heat bath can still reach the aimed temperature and a temperature gradient can be established in the middle region. With the increase of τ from 0.01, the temperature gradient in the middle region is almost the same, but the mean temperature decreases until τ=0.2. This results in a decrease in heat flux when τ<0.2. Particularly when τ=0.04, TJ at the left boundary is nearly zero, while TJ at the right boundary is about 5K. This should be a numerical artifact because TJ has its physical origin from the thermal interface resistance between heat bath and the middle region, thus cannot be eliminated by tuning parameter. With further increase of τ (0.2<τ<300), temperature profile becomes correct and stable:

both $J_L$ and temperature gradient become insensitive to τ, thus the calculated thermal conductivity in Fig. 3(d) is almost a constant with small fluctuation. Next, when $\tau > 300$, temperature gradient becomes smaller and $J_L$ decreases quickly. This is because in the large $\tau$ range, the distribution of ζ values becomes a δ-function [21]. As a result, it would require longer simulation times in order to ensure the decay of correlations. This causes the heat bath cannot reach the aimed temperature due to the limited simulation time, also a smaller temperature gradient. Therefore, from the consideration of a practical computational time, one should not choose too large τ.

For Langevin heat bath, in the weak coupling limit $\lambda \rightarrow 0$ (e.g. λ=0.1 in Fig. 3(b)), the heat bath cannot reach the aimed temperature. However, in the strong coupling limit $\lambda \rightarrow \infty$ (e.g. λ=500 in Fig. 3(b)), although the heat baths can reach the aimed temperature, large TJ is observed at the boundary. In both cases, small temperature gradient is generated, which induces a small heat flux. In the middle range of λ, a temperature profile with correct heat bath temperature and small TJ can be established. This causes $J_L$ and thermal conductivity first increase then decrease with the increase of λ as shown in Fig. 3(c) and (d). All these are consistent with the results obtained from Fermi-Pasta-Ulam (FPU) chain in Ref. [21].

In MD calculation, it must be confirmed that the temperature profile is correct and a temperature gradient can be well established in order to achieve accurate prediction results. Based on these considerations and the results shown in Figure. 3, an intermediate value of λ (from 1 to 100) is recommended for Langevin heat bath. For NH heat bath, fig. 3 suggests τ from 0.2 to 300 is the optimal choice in numerical simulation when NH heat bath is applied. Moreover, we can see from Fig. 3(d) that the discrepancy of thermal conductivity in Fig. 2(e) is mainly due to the choice of heat bath parameter: the same result can be obtained if the parameter is chosen properly.

Heterogeneous materials such as graphene based [6] and carbon nanotube based [7] nano-junctions are promising candidates for thermal rectifier application. In these systems, it is problematic to define the thermal conductivity according to Fourier's law, due to the large temperature jump at the interface of two different materials. Instead, because of the asymmetry of the heterogeneous materials, people are more interested in the rectification effect of heat current in such materials [32], namely the difference between the heat current in different directions. In the following part, we extend our study to heterogeneous materials and discuss the effect of heat bath parameter on the heat flux rectification.

We use Si/Ge NW junction as an example. It has a fixed cross section of $3\times 3$ unit cells, 5 unit cells of Si and 5 unit cells of Ge in the longitudinal direction, and 3 layers of heat bath are applied at each end. Here we define $J_+$ ($J_-$) to be the heat current of non-equilibrium steady state when Si (Ge) end is attached to the high temperature heat bath (310K) and temperature of the other heat bath is 290K. We define the rectification ratio to be:

$$RE = (J_+ - J_-)/J_- \quad . \tag{5}$$

Figure 4(a), (b) shows the dependence of heat current on heat bath parameters. It has been checked that temperature profiles are all correct for these parameters for both NH and Langevin heat bath used. With NH heat bath, $J_+$ and $J_-$ have the distinct dependence on $\tau$. In the small $\tau$ limit, $J_+$ is much larger than $J_-$. With the increase of $\tau$, $J_+$ drops rapidly and finally converges to a small value, while $J_-$ first increases, then decreases, and finally converges to a value which is slightly larger than $J_+$. As a result, there exists a large nontrivial $RE$ in the small $\tau$ limit as shown in Fig. 4(c) and $RE$ changes from positive to negative when $0.05 \leq \tau \leq 100$. Thus NH heat bath fails to give a consistent result. However, with Langevin heat bath, Fig. 4(b) shows the same

dependence on λ for both $J_+$ and $J_-$, and there is only a small difference between them. This results a small value of *RE* as shown in Fig. 4(c) for Langevin heat bath. More importantly, Langevin heat bath can produce a consistent result ( $RE<0$ ) that $J_-$ is always larger than $J_+$, regardless the heat bath parameter λ. Si-Ge nanowire is one graded massed nano-junction. The results calculated with Langevin heat bath and NH heat bath with large τ ( $\tau>1$ ) suggest that the heat flux runs preferentially along the direction of decreasing mass. This conclusion is consistent with experimental result in Ref. [4]. In Ref. [4], Chang *et al.* demonstrated the thermal rectification effect in carbon and boron nitride nanotubes which were mass-loaded inhomogeneously with heavy molecules. A larger heat flow was observed when heavy-mass end is at higher temperature.

IV. CONCLUSIONS

In conclusion, we have studied the impacts of heat bath applied in molecular dynamics simulations of nano structure thermal properties. Due to the existence of localized edge modes and accumulation of LEM because of the deterministic characteristic of Nosé-Hoover heat bath, multiple layers of NH heat bath are required in order to reduce the temperature jump at the boundary. Even with one layer of Langevin heat bath, it can prevent the accumulation of LEM due to its stochastic excitation of all modes, giving rise to a small temperature jump at the boundary. In addition, in order to obtain the correct temperature profile, intermediate values, $0.2<\tau<300$ for NH heat bath and $1<\lambda\leq100$ for Langevin heat bath are recommended. Moreover, in the study of heat current rectification in heterogeneous materials, Langevin heat bath is recommended because it can produce consistent results with experiment, regardless the heat bath parameter used. As our simulations are based on underlying physics, our conclusion is quite general and can be applied to other materials.

**ACKNOWLEDGEMENTS**

J. Chen would like to thank J. W. Jiang for discussions. This work is supported in part by an ARF grant, R-144-000-203-112, from the Ministry of Education of the Republic of Singapore, and Grant R-144-000-222-646 from National University of Singapore

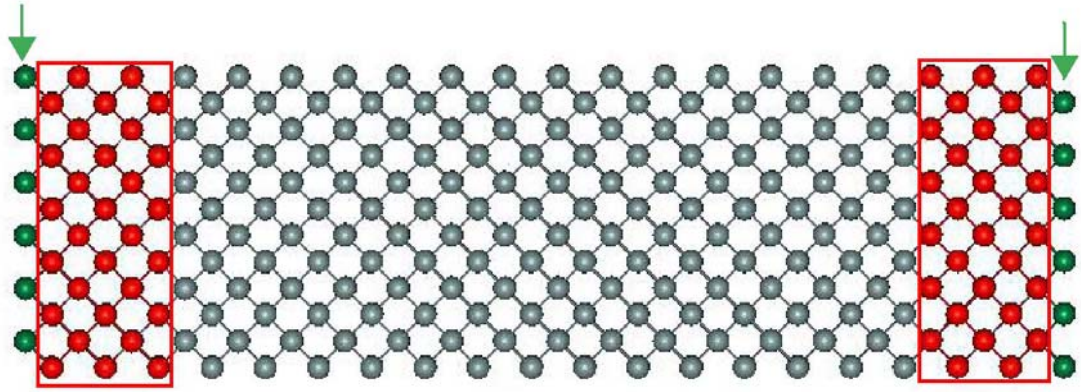

Figure 1. (color online) Schematic picture of the SiNWs. Heat Bath layers are in the rectangular box. The arrows pinpoint the boundary layers.

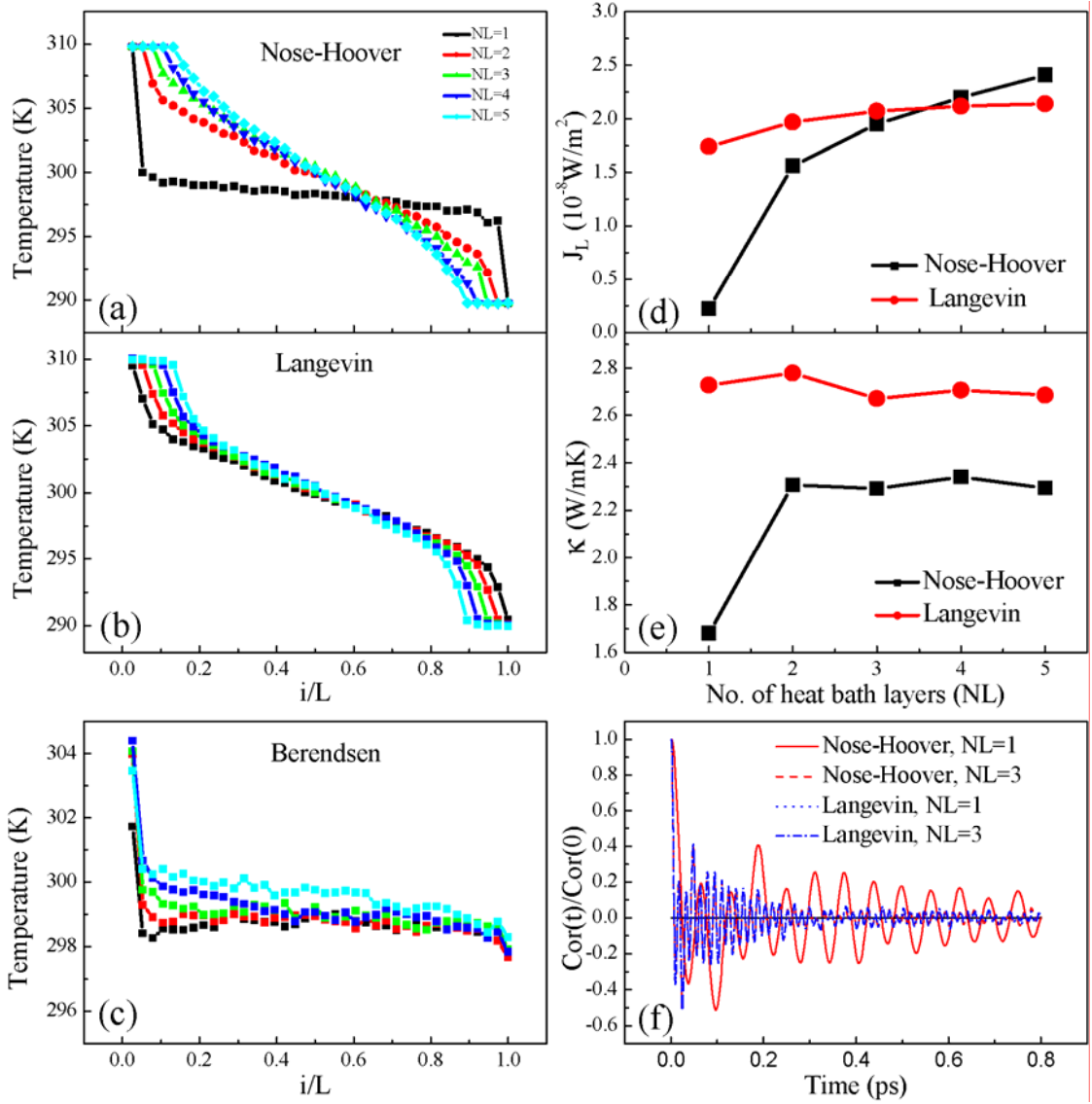

Figure 2. (color online) Impacts of the number of heat bath layers (NL) on thermal properties of SiNWs. (a) Temperature profile with different NL of Nosé-Hoover heat bath. (b) Temperature profile with different NL of Langevin heat bath. (c) Temperature profile with different NL of Berendsen heat bath. (d) Heat current $J_L$ along the longitudinal direction versus NL. (e) Thermal conductivity $\kappa$ versus NL. (f) Normalized autocorrelation function of velocity in different heat baths. Except for the solid line, the other three lines almost overlap with each other.

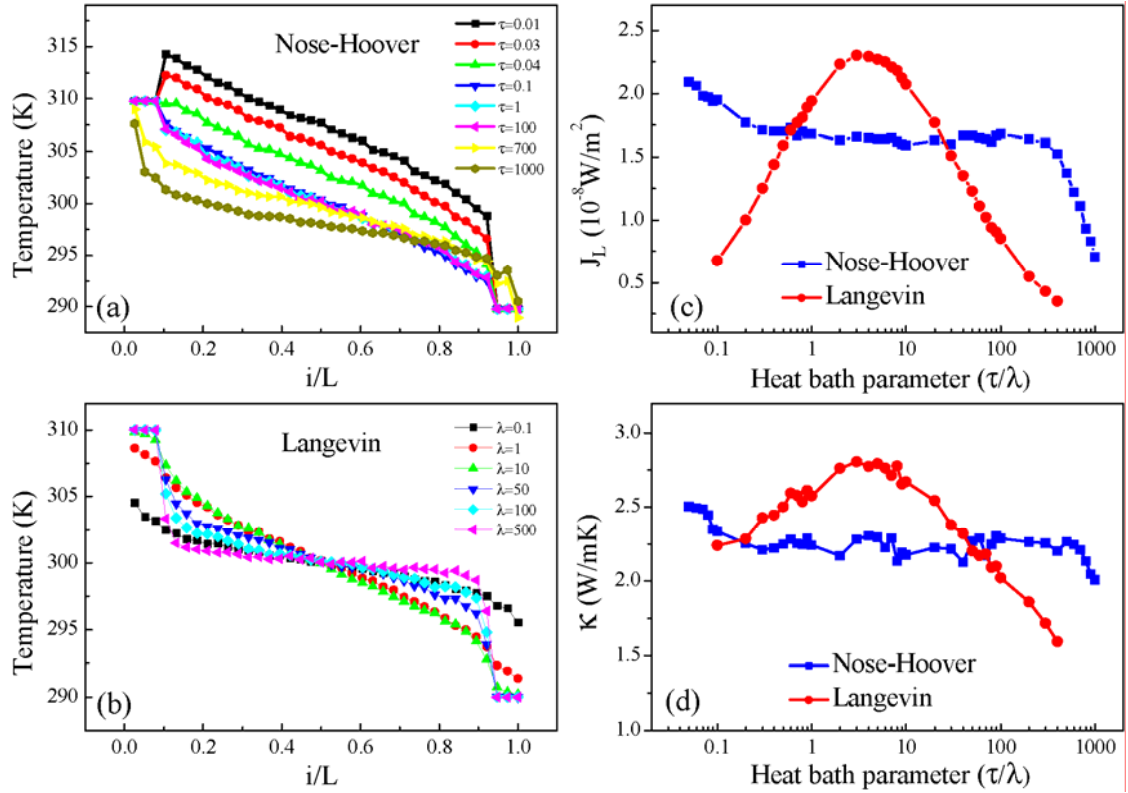

Figure 3. (color online) Impacts of heat bath parameter on thermal properties of SiNWs. The NWs has a fixed cross section of 3×3 unit cells. In the longitudinal direction, it has a fixed length of 10 unit cells with 3 layers of heat bath at each end. (a) Temperature profile with parameter $\tau$ of Nosé-Hoover heat bath. (b) Temperature profile with parameter $\lambda$ of Langevin heat bath. (c) Heat current $J_L$ versus $\tau/\lambda$ for Nosé-Hoover/ Langevin heat bath. (d) Thermal conductivity $\kappa$ versus $\tau/\lambda$ for Nosé-Hoover/ Langevin heat bath. Here 0.05≤τ≤1000, and 0.1≤λ≤400.

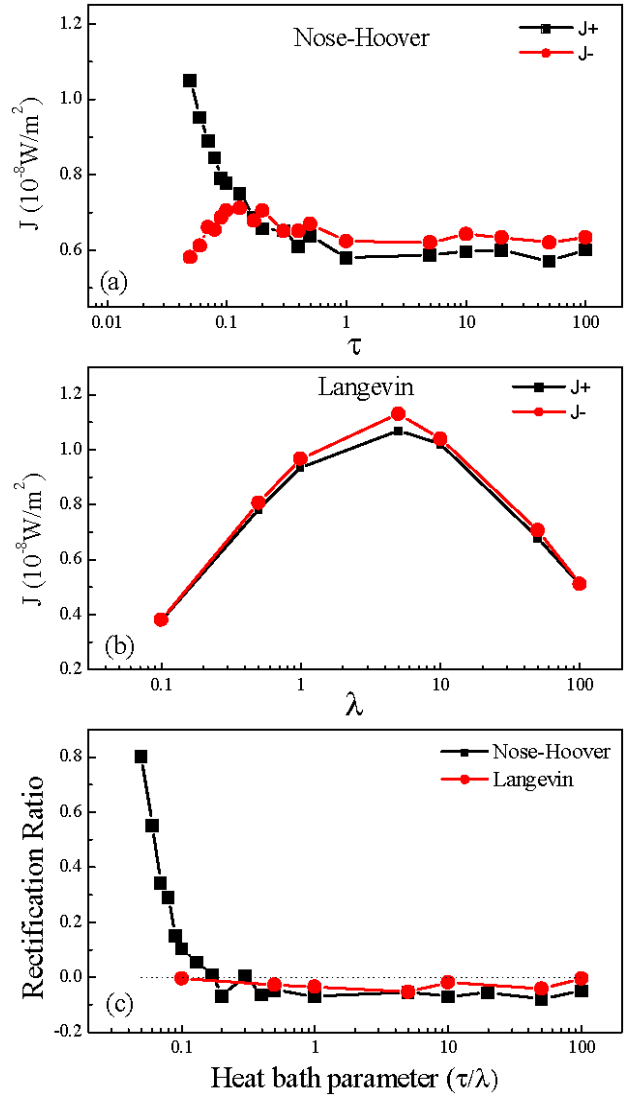

Figure 4. (color online) $J_\pm$ in Si/Ge NWs versus heat bath parameters. (a) $J_\pm$ versus parameter $\tau$ of Nosé-Hoover heat bath. (b) $J_\pm$ versus parameter $\lambda$ of Langevin heat bath. (c) Rectification ratio versus heat bath parameter. The square and circle denote the results for Nosé-Hoover($\tau$) and Langevin($\lambda$) heat bath, respectively.